\documentclass[fleqn,usenatbib]{mnras}

\usepackage{newtxtext,newtxmath}

\usepackage[T1]{fontenc}
\usepackage{ae,aecompl}
\usepackage{siunitx}
\usepackage[para]{threeparttable}

\usepackage{graphicx}	
\usepackage{amsmath}	
\usepackage{wrapfig,lipsum}
\usepackage{hyperref}
\usepackage{upgreek}

\title[FRB search]{A targeted search for repeating fast radio bursts with the MWA}

\author[J. Tian et al.]{J. Tian$^{1}$\thanks{E-mail: jun.tian@postgrad.curtin.edu.au}, G. E. Anderson$^1$, P. J. Hancock$^{1,2}$, J. C. A. Miller-Jones$^1$, M. Sokolowski$^1$,
\newauthor C. W. James$^1$, N. D. R. Bhat$^1$, N. A. Swainston$^1$, D. Ung$^1$, B. W. Meyers$^{1,3}$
\\
$^1$International Centre for Radio Astronomy Research, Curtin University, GPO Box U1987, Perth, WA 6845, Australia\\
$^2$Curtin Institute for Computation, Curtin University, GPO Box U1987, Perth, 6845, WA, Australia\\
$^3$Department of Physics and Astronomy, University of British Columbia, 6224 Agricultural Road, Vancouver, BC V6T 1Z1 Canada
}%

\date{Accepted XXX. Received YYY; in original form ZZZ}

\pubyear{2022}

\begin{document}
\label{firstpage}
\pagerange{\pageref{firstpage}--\pageref{lastpage}}
\maketitle

\begin{abstract}
We present a targeted search for low-frequency (144--215\,MHz) FRB emission from five repeating FRBs using 23.3\,hr of archival data taken with the Murchison Widefield Array (MWA) Voltage Capture System (VCS) between 2014 September and 2020 May. This is the first time that the MWA VCS has been used to search for FRB signals from known repeaters, which enables much more sensitive FRB searches than previously performed with the standard MWA correlator mode. We performed a standard single pulse search with a temporal and spectral resolution of $400\,\upmu$s and 10\,kHz, respectively, over a $100\,\text{pc}\,\text{cm}^{-3}$ dispersion measure (DM) range centred at the known DM of each studied repeating FRB. No FRBs exceeding a $6\sigma$ threshold were detected. The fluence upper limits in the range of 32--1175\,Jy\,ms and 36--488\,Jy\,ms derived from 10 observations of FRB 20190711A and four observations of FRB 20201124A respectively, allow us to constrain the spectral indices of their bursts to $\gtrsim-1$ if these two repeaters were active during the MWA observations. If free-free absorption is responsible for our non-detection, we can constrain the size of the absorbing medium in terms of the electron temperature $T$ to $<1.00\times(T/10^4\text{K})^{-1.35}\,\text{pc}$, $<0.92\times(T/10^4\text{K})^{-1.35}\,\text{pc}$ and $<[0.22\text{--}2.50]\times(T/10^4\text{K})^{-1.35}\,\text{pc}$ for FRB 20190117A, 20190711A, and 20201124A, respectively. However, given that the activities of these repeaters are not well characterised, our non-detections could also suggest they were inactive during the MWA observations.
\end{abstract}




\section{INTRODUCTION }

Fast radio bursts (FRBs) are impulsive bursts of radio emission with durations ranging from microseconds to milliseconds (e.g. \citealt{Cho20, Nimmo21}) and fluences from $\sim0.01$ to 1000\,Jy\,ms (e.g. \citealt{Petroff16, Shannon18}). Their typically large excess dispersion measures (DMs) with respect to the Galactic contribution in the line of sight suggests an extra-galactic origin. This has been confirmed through identifying the host galaxies of a few sub-arcsecond localised FRBs (e.g. \citealt{Chatterjee17, JP20}). Although dozens of models have been proposed to explain the emission behavior of FRBs (see \citealt{Zhang20c} for a review), their physical origin remains unknown.

Over the last couple years, the FRB population has rapidly expanded, with more than 700 FRBs published so far, mostly thanks to the Canadian Hydrogen Intensity Mapping Experiment (CHIME; \citealt{CHIME22}) Fast Radio Burst project (CHIME/FRB; \citealt{CHIME18}). In the FRB population there are apparently two classes of FRBs: repeaters \citep{Spitler16} and non-repeaters \citep{Petroff15, Shannon18}, and their progenitors and/or emission mechanisms could potentially be different (e.g., \citealt{CHIME19, Fonseca20}). The existence of two populations of FRBs is further supported by recent studies using a large sample of CHIME FRBs \citep{CHIME21, Pleunis21b}. Currently, there are 24 sources known to repeat \citep{Spitler16, CHIME19, CHIME19b, Kumar19, Fonseca20, CHIME21}, and there is evidence for periodic activity from two of them, FRB 20121102A \citep{Rajwade20, Cruces21} and 20180916B \citep{CHIME20b}, which might reflect an orbital \citep{Ioka20}, rotational \citep{Beniamini20} or precession \citep{Levin20} period. The recent discovery of unusually bright FRB-like emission from the Galactic magnetar SGR 1935+2154 suggests that magnetars could produce at least some of the extra-galactic FRBs \citep{Bochenek20, CHIME20}. This association is further supported by the latest identification of sub-second periodicity from the sub-components of FRB 20191221A \citep{CHIME21d}.

Repeating and non-repeating FRBs often show different features in their burst properties. For example, on average, repeating FRBs tend to have larger burst widths than non-repeating FRBs \citep{Scholz16, CHIME19b, Fonseca20, Pleunis21b}. The downward drifting of subpulses in frequency with time, i.e. the ''sad trombone'' effect, appears to be a common feature among repeating FRBs \citep{Hessels19, CHIME19c, Pleunis21b}. While some repeating FRBs share flat polarisation position angles within and between bursts, such as FRB 20121102A and FRB 20180916B \citep{Michilli18, Nimmo21}, the bursts of FRB 20180301 show a diversity of position angle swings \citep{Luo20}. Detailed spectrotemporal and polarimetric studies of more repeating FRBs are needed to confirm and explain these features.

Up until very recently, 
FRB repeating sources have been 
detected at frequencies between 300\,MHz \citep{Chawla20, Pilia20, Parent20} and 
8\,GHz \citep{Gajjar18}.
However, simultaneous observations of repeating FRBs in different bands suggest
that individual pulses are narrow-band, yet are scattered across a wide range of frequencies with time
\citep{Gourdji19}. For example, FRB 20180916B has been detected up to 5.3\,GHz \citep{Bethapudi22}, yet
the absence of simultaneous detections at frequencies 2.3 and 8.4\,GHz demonstrates its frequency dependent activity \citep{Pearlman20}. While \citet{Chawla20} detected FRB 20180916B at $300\text{--}400$\,MHz using the Green Bank Telescope, no emission was seen contemporaneously at $110\text{--}188$\,MHz using the Low Frequency Array (LOFAR). Such narrow-band emission has also been demonstrated for FRB 20121102A \citep{Gourdji19, Majid20} and FRB 20190711A \citep{Kumar21c}.

To date, there have been many searches for FRB emission below 300\,MHz, including simultaneous, multiband, targeted \citep{Law17, Sokolowski18, Houben19} and wide-field blind \citep{Coenen14, Karastergiou15, Tingay15, Rowlinson16, Sanidas19} searches using LOFAR, the Murchison Wide-field Array (MWA) and the Long-wavelength Array.
In addition, searches for prompt and dispersed (FRB-like) signals predicted to be associated with gamma-ray bursts have also been conducted with MWA and LOFAR via rapid-response observations \citep{Kaplan15, Rowlinson19b, Rowlinson21, Anderson20, Tian22a, Tian22b},
none of which have yielded a detection.
However, recently 
\citet{Pleunis21} reported the detection of 18 bursts from FRB 20180916B between $110\text{--}188$\,MHz using LOFAR, the only detections of any FRB below 300\,MHz to date. This confirms the existence and the detectability of low frequency bursts from repeating FRBs at cosmological distances, which are not limited by propagation effects or the FRB emission mechanism. 
Additionally,
this same repeater appears to undergo chromatic periodic activity where its activity window is wider and occurs later with decreasing frequency.
This is demonstrated by the $\sim3$ day delay in the peak activity between bursts observed by CHIME/FRB at 600\,MHz and those observed by LOFAR at 150\,MHz \citep{Pleunis21, Pastor21}.

There are several reasons for the dearth of FRB detections at low frequencies. First is an increase in the sky background temperature at low radio frequencies (e.g. \citealt{Costa08, Cong21}). This can increase the noise level in low frequency observations and reduce the observed signal-to-noise ratio (S/N). As FRB signals propagate through intervening ionized media before reaching the Earth, the effects of scatter broadening of the pulse profiles and intrachannel dispersive smearing, which are more significant at low frequencies, can also reduce the peak S/N. 
In addition, 
repeating FRBs are also known to suddenly turn on and enter periods of long-term
($>$ a few years) activity 
(e.g., FRB 20201124A; \citealt{Lanman22}). 
This makes it difficult to define 
proper burst rates for sources that 
exhibit burst clustering, which may be
more/less active at higher/lower 
frequencies (e.g., FRB 20121102A and FRB 20180916B; \citealt{Josephy19, Pearlman20}).
All of these present challenges to FRB searches at low radio frequencies.

Nonetheless, low frequency FRB searches are very important. A real detection at low frequencies would complement high frequency detections for broadband measurements of the FRB spectrum, allowing more reliable studies of the burst energetics and the emission mechanism. It would also allow us to better constrain the local environments of FRBs based on propagation effects such as free-free absorption. As this effect is more obvious at lower frequencies 
(optical depth scales as  $\tau_\text{ff}\propto\nu^{-2.1}$)
, FRB measurements at low frequencies would place more stringent constraints on the size of the emission site \citep{Pleunis21}. 
Low frequency FRB emission is also sensitive to other propagation effects such as the dispersion, scattering and Faraday rotation, and could provide precise measurements of the dispersion measure, scattering timescale and rotation measure of FRBs \citep{Petroff22}.
Therefore, low frequency FRB searches are well motivated scientifically in spite of the observational challenges.

In this paper, we use the archival MWA observations taken with the voltage capture system (VCS; \citealt{Tremblay15}), which has a high temporal resolution of $100\,\upmu$s, to search for low frequency (144--215\,MHz) FRB signals from known repeating FRBs. In Section 2, we describe our selection of MWA observations in the data archive and the data processing and analysis we used for the FRB search. Our results are then presented in Section 3. We discuss the implications of our results for FRB sources and emission models in Section 4, followed by conclusions in Section 5.

\section{Observations and analysis}

\subsection{MWA VCS mode}

The MWA is a low frequency radio telescope with an operational frequency range of $80\text{--}300$\,MHz, an instantaneous bandwidth of 30.72\,MHz and a field of view ranging from $\sim300\text{--}1000\,\text{deg}^2$ \citep{Tingay13, Wayth18}. 
There are two observing modes of the MWA: the standard correlator with a time and frequency resolution of 0.5\,s/10\,kHz and the VCS with a resolution of $100\,\upmu$s/10\,kHz. We chose to inspect MWA VCS observations for FRBs from known repeaters due to 
their high time resolution, making them specifically sensitive to narrow ($\sim$\,ms) pulsed and therefore dispersed signals, i.e. FRB emission.

The MWA VCS mode has been extensively used for pulsar studies and searches \citep{Bhat16, Xue17, McSweeney17, Meyers18, Swainston21}. In September 2018, an all sky pulsar search project, the Southern-sky MWA Rapid Two-meter (SMART) survey \citep{Bhat22}, was commenced to search the entire sky south of $+30^{\circ}$ in declination for pulsars, with regular data collection planned until $\sim2023$. 
This survey, along with targeted observations toward a number of already known pulsars, generates a large amount of VCS observations in the MWA archive.
These all-sky archival data at the highest time resolution available with the MWA, combined with the MWA's large field of view
are an invaluable resource to exploit for 
transient studies and enable the search for low frequency bursts from known repeating FRB sources. The observations we selected for FRB searches were taken between 2014 September and 2020 May, spanning the MWA Phase I and Phase II, which differ in array configuration, and baseline length and distribution. The MWA Phase I presents 
an angular resolution of $\sim2$\,arcmin at 185\,MHz \citep{Tingay13}. The MWA phase II has two configurations: extended and compact configurations with angular resolutions of $\sim1$ and $\sim10$\,arcmin, respectively \citep{Wayth18}. Both can be used for FRB searches.

\subsection{FRB selection}\label{sec:FRB_selection}

\begin{table*}
\begin{threeparttable}
\centering
\resizebox{2\columnwidth}{!}{\hspace{-0cm}\begin{tabular}{c c c c c c c c c}
\hline
 FRB & Instrument\tnote{1} & RA & Dec & Frequency\tnote{2} & Burst rate\tnote{3} & Fluence\tnote{4} & Burst Width\tnote{4} & DM\tnote{5} \\
 & & (deg) & (deg) &  & ($\text{hr}^{-1}$) & (Jy\,ms) & (ms) & ($\text{pc}\,\text{cm}^{-3}$) \\
\hline
20190116A\tnote{a} & CHIME & $192.33\pm0.15$ & $27.15\pm0.24$ & 400\text{--}700\,MHz & 0.25 & 0.8\text{--}2.8 & 1.5\text{--}4 & 441  \\
20190117A\tnote{b} & CHIME & $331.71\pm0.15$ & $17.37\pm0.26$ & 400\text{--}800\,MHz & 0.26 & 5.0\text{--}12 & 0.64\text{--}5.2 & 393.6  \\
20190213A\tnote{b} & CHIME & $31.72\pm0.25$ & $20.08\pm0.30$ & 400\text{--}600\,MHz & 0.12 & 0.6\text{--}3.0 & 4\text{--}10 & 651.45   \\
20190711A\tnote{c} & ASKAP & 329.4195 & -80.358 & 1.1\text{--}1.4\,GHz & --\tnote{6} & 1.4\text{--}34 & 1.0\text{--}6.5 & 593.1  \\
20201124A\tnote{d} & CHIME & $76.99\pm0.52$ & $26.19\pm0.53$ & 550--750\,MHz & 16\tnote{7} & 2.6--108 & 6.1--59.3 & 411  \\
\hline
\end{tabular}}
\caption{Details of the five known repeaters that can be observed by the MWA. References for the repeater properties: a: \citet{CHIME19b}; b: \citet{CHIME20c}; c: \citet{JP20, Kumar21c}; d: \citet{Kumar21a, Marthi21, Piro21, Lanman22}.\\
1: The radio instrument that discovered and localised each repeater.\\
2: The frequency of detected emission from each repeater.\\
3: The burst rate inferred by the ratio between the number of detected bursts and the total exposure time of each repeater. The burst rates are all based on CHIME observations except for FRB 20190711A and 20201124A (see the notes below). An estimation of expected bursts during the MWA observations is given in Table~\ref{Results} assuming the burst rate is constant.\\
4: The range of fluences and widths observed for each repeater by CHIME or ASKAP (see Section~\ref{sec:FRB_selection}).\\
5: The known DM reported for each repeater.\\
6: The burst rate of FRB 20190711A is extremely low, considering $\sim300$\,hr of follow up observations using different telescopes identified only one repeat burst \citep{Kumar21c}.\\
7: The burst rate of FRB 20201124A is estimated for the period of high activity after 2021 March, while its rate prior to discovery (based on the non-detection over the pre-discovery total observed time) is as low as $<3.4\,\text{day}^{-1}$ \citep{Marthi21, Piro21, Lanman22}.
}
\label{FRBs}
\end{threeparttable}
\end{table*}

We used the first complete FRB catalogue\footnote{\url{https://www.herta-experiment.org/frbstats/}} (including FRB events published in the Transient Name Server\footnote{\url{https://www.wis-tns.org/}}, FRBCAT and the CHIME/FRB Catalogue; \citealt{Petroff16, CHIME21}) to identify targets for our FRB search, i.e. those repeaters that could be viewed 
by the MWA.
Since the majority of repeaters have been detected by CHIME, which looks at the Northern sky (CHIME/FRB is sensitive to sky locations with declinations $>-11^\circ$; \citealt{CHIME21}), only a few of them are located in the MWA observable sky.
We filtered through the population of known repeating FRBs using the criterion of their maximum elevation above the MWA's horizon of $>30^\circ$ (for the choice of the elevation limit, see \citealt{Hancock19}), and obtained a list of five repeating FRBs (see Table~\ref{FRBs}). 

The five repeating FRBs display varying burst widths, fluences and degrees of activity and repetition rates, as shown in Table~\ref{FRBs}. 
FRB 20190711A was first detected by ASKAP in the frequency range 1.1--1.3\,GHz \citep{JP20}, and hundreds of hours of follow-up observations with ASKAP and Parkes detected only one repeat burst at 1.4\,GHz in the Parkes observations \citep{Kumar21c}, indicating an extremely low repetition rate. We do not report a burst rate for this repeater in Table~\ref{FRBs}.
FRB 20190116A, 20190117A, 20190213A and 20201124A were first detected by CHIME in the frequency range $400\text{--}800$\,MHz, with a burst rate simply estimated by the ratio of the number of detected bursts to the exposure time of the CHIME system on the source \citep{CHIME19b, CHIME20c}.
The brightest repeater in this sample, FRB 20201124A, was subsequently detected by multiple instruments including the Karl G. Jansky Very Large Array (VLA) at 1.5\,GHz \citep{Law21}, the Australian Square Kilometre Array Pathfinder (ASKAP) at 864.5\,MHz \citep{Kumar21a, Kumar21b} and the Five-hundred-meter Aperture Spherical radio Telescope (FAST) at 1.2\,GHz \citep{Xu21}, and reported to enter a sudden period of high activity in 2021 March \citep{Marthi21, Piro21, Lanman22}. Thanks to these radio observations between 864.5\,MHz and 1.5\,GHz, FRB 20201124A is the only repeater in our sample that has a measured spectral index $\alpha=-5.82^{+0.68}_{-0.84}$ \citep{Kumar21f}.
For the CHIME repeaters, the burst rates quoted in Table~\ref{FRBs} assume an average rate that does not take into account the likely variable nature of a potential activity window, and thus may not apply to the time windows of the MWA observations inspected in our analysis (see discussions in Section~\ref{sec:rate}). Note that the CHIME rates in Table~\ref{FRBs} are all based on CHIME observations except for FRB 20201124A, which is estimated for the period of high activity after 2021 March while its rate prior to discovery is as low as $<3.4\,\text{day}^{-1}$ \citep{Marthi21, Piro21, Lanman22}.

We used the MWA All-Sky Virtual Observatory\footnote{\url{https://asvo.mwatelescope.org/}} (ASVO) to search for all VCS observations that overlap with the positional errors of the five repeating FRBs, and found 61 observations in the MWA data archive.
We then estimated the sensitivity of each VCS observation in the directions of the five repeating FRBs based on the Full Embedded Element model of the MWA \citep{Sokolowski17}. We selected only those observations with at least 20\% of the maximum sensitivity in the primary beam. This resulted in a total of 25 observations with integrations between 15\,min and 1.5\,hr (23.3\,hr in total) and central frequencies between 144\,MHz and 215\,MHz, as listed in Table~\ref{MWAObs}. Of these observations, 12 were taken in the MWA Phase II compact configuration, and the others were recorded in the higher angular resolution of Phase I and the Phase II extended configuration.

\begin{table*}
\begin{threeparttable}
\centering
\resizebox{2.2\columnwidth}{!}{\hspace{-0.5cm}\begin{tabular}{c c c c c c c c c c c c c}
\hline
 FRB & Date\tnote{1} & Start time\tnote{1} &Obs. ID\tnote{1} & Config.\tnote{1} & Dur.\tnote{2} & Freq.\tnote{3} & Elev.\tnote{4} & Cal.\tnote{5} & Beams\tnote{6} & Min. detectable  & Candidates & Candidates\\
 & &(UT) && & (s) & (MHz) & (deg) & &  & flux density\tnote{7}  & above $6\sigma$\tnote{8}&with friends\tnote{8} \\
 & & && & & && & & (Jy)  & & \\
\hline
20190116A & 2018-04-05 &15:27:58&1206977296 & IIE & 3600 & 185 & 36.1 & HydA & 550 & 29.36 & 40368&26\\
        & 2018-03-27 &15:49:58&1206201016& IIE & 3600 & 185 & 35.9 & HydA & 550 & 27.20 &42693&194\\
\hline 
20190117A & 2015-10-13 &11:49:27&1128772184& I & 3600 & 185 & 45.0 & PicA & 540 & 29.48 &41633 &51\\
        & 2017-08-01 &17:41:58&1185644536& IIC & 3600 & 144 & 44.0 & 3C444 & 2 & 25.70 & 137 &0\\
        & 2017-08-08 &17:19:58&1186248016& IIC & 3600 & 185 & 44.8 & PicA & 2 & 22.43 & 169 &0\\
        & 2017-11-18 &11:01:58&1195038136& IIE & 3600 & 185 & 43.3 & 3C444 & 540 & 23.25 & 41266 &44\\
        & 2018-09-24 &13:51:02&1221832280& IIC & 4800 & 154 & 45.4 & 3C444 & 2 & 24.82 & 214 &0\\
        & 2018-10-08 &14:01:02&1223042480& IIC & 4800 & 154 & 40.8 & PicA & 2 & 26.46 & 208 &0\\
\hline 
20190213A & 2016-11-22 &12:35:03&1163853320& IIC & 4800 & 185 & 42.8 & 3C444 & 6 & 24.80 & 2611 &17\\
        & 2018-11-01 &14:37:02&1225118240& IIC & 4800 & 154 & 42.8 & PicA & 6 & 26.15 & 619&0\\
        & 2019-09-10 &19:08:46&1252177744& IIC & 4800 & 154 & 41.8 & PicA & 6 & 33.19 & 1024 &0\\
\hline
20190711A & 2014-09-23 &11:14:56&1095506112& I & 3600 & 185 & 34.5 & PicA & 1 & 45.01 &72 &0\\
        & 2015-07-01 &21:38:15&1119821912& I & 2400 & 185 & 34.4 & 3C444 & 1 & 25.62 & 54 &0\\
        & 2015-10-13 &11:49:27&1128772184& I & 3600 & 185 & 36.3 & PicA & 1 & 29.50 & 103 &0\\
        & 2016-09-16 &11:36:31&1158061008& IIC & 1440 & 172 & 33.9 & HerA & 1 & 52.68 & 178 &0\\
        & 2017-08-08 &17:19:58&1186248016& IIC & 3600 & 185 & 36.2 & PicA & 1 & 22.13 & 75 & 0\\
        & 2017-11-18 &11:01:58&1195038136& IIE & 3600 & 185 & 35.9 & 3C444 & 1 & 23.25 & 965&0\\
        & 2018-11-23 &12:05:58&1227009976& IIC & 4800 & 154 & 33.9 & 3C444 & 1 & 27.89 & 115 &9\\
        & 2019-07-25 &14:46:14&1248101192& IIE & 5400 & 215 & 34.2 & PicA & 1 & 55.86 & 856 &0\\
        & 2020-04-29 &21:02:28&1272229366& IIE & 1500 & 167 & 34.4 & 3C444 & 1 & 42.67 &38 &0\\
        & 2020-05-02 &21:01:58&1272488536& IIE & 1500 & 167 & 34.7 & 3C444 & 1 & 40.35 & 31&0\\
\hline 
20201124A & 2014-11-07 &16:53:20&1099414416& I & 1200 & 185 &   34.3   & HydA & 576 & 18.41 & 54021 &2\\
        & 2015-10-03 &20:29:11&1127939368& I & 3600 & 157 &    36.7  & PicA & 576 & 21.15 & 41171&17\\
        & 2016-10-14 &20:01:19&1160510496& IIC & 1200 & 185 &  36.8  & PicA & 16 & 21.07 & 3524 &183\\
        & 2016-12-08 &15:39:59&1165246816& IIC & 900 & 154 &    36.9 & HydA & 16 & 22.00 & 293 &0\\
\hline
\end{tabular}}
\caption{MWA observations used for the FRB search and the corresponding results.\\
1: The MWA observation date and start time in UT, observation ID and array configuration, including phase I ('I'), or phase II extended ('IIE') or compact ('IIC');\\
2: The duration of each MWA observation;\\
3: The central frequency of each MWA observation;\\
4: The maximum elevation of the FRB in the MWA's field of view during the observation;\\
5: The calibrator used to calibrate the MWA observations;\\
6: The number of synthesised beams within the positional error of the FRB;\\
7: The minimum detectable flux density calculated using the radiometer equation for each observation (see Section~\ref{sec:sensi});\\
8: The number of candidates above $6\sigma$ resulting from the single pulse search on each observation and those passing the friends-of-friends algorithm (see Section~\ref{sec:search}).
}
\label{MWAObs}
\end{threeparttable}
\end{table*}

\subsection{Data processing}

For the data processing we used the VCS data processing pipeline, which was initially developed for pulsar detections (e.g. \citealt{Bhat16}, \citealt{McSweeney17}, \citealt{Meyers17} and \citealt{Ord19}). It automates the reduction of MWA VCS data of targeted sources, including downloading, calibration and beamforming at the target positions. The final data product is a time series of Stokes parameters written into the \mbox{PSRFITS} format \citep{Hotan04}, which can be further analyzed by the {\sc presto} software package\footnote{\mbox{\url{https://github.com/scottransom/presto}}} \citep{Ransom01}. Here we present specific details regarding calibrating the VCS data and beamforming at the positions of the target FRBs.

\subsubsection{Calibration}\label{sec:cal}

For each MWA observation, we need to determine the direction independent complex gains, including amplitudes and phases, for each constituent tile ($4\times4$ dipole array) through the calibration process (for details see \citealt{Ord19}). 
We selected a bright source that had been observed in the standard correlator mode within 12 hours of the VCS observation as the calibrator source.
The name of the calibrator source used to calibrate each VCS observation is listed in Table~\ref{MWAObs}.
The Real Time System (RTS; \citealt{Mitchell08}) was used to generate a calibration solution for the amplitude and phase for each of the $24\times1.28$\,MHz sub-bands and each tile from the visibilities. We inspected these solutions and discarded tiles showing poor calibration solutions. We also excised the edge channels (0\textendash7 and 120\textendash127) of each of the 24 sub-bands to alleviate the aliasing effects resulting from the channelization process. We applied this calibration process to each MWA observation before coherently summing the power from the constituent tiles.

\subsubsection{Coherent beamforming}\label{sec:beamform}

In order to maximize our sensitivity to any millisecond-duration signals from the five repeating FRBs, we coherently summed the voltages from individual MWA tiles to form a tied-array beam in the FRB directions (i.e., coherent beamforming; \citealt{Ord19, Swainston22}).
This can potentially gain more than an order of magnitude increase in sensitivity for each phase centered beam over
incoherent beamforming, which simply sums up the power from each tile to preserve a large field of view (e.g. \citealt{Bhat16}).
The performance of coherent beamforming is affected by a few factors, such as the quality of the calibration solution and the pointing direction of the telescope.

Here we briefly summarise the beamforming process (for more details see \citealt{Swainston22}). The essential step is converting cable and geometric delays to the pointing center into phase shifts for each tile. With the knowledge of the delay model and the complex gain information from the calibration solution derived in the calibration process, we can obtain the tile based gain solution to equalise the tile gains and phase all tiles to the same direction. 

We used the coherent beamforming to phase all MWA tiles to the known positions of the five repeating FRBs, as listed in Table~\ref{FRBs}. 
The number of coherent beams we needed to form for each FRB depends on the positional error of the FRB and the angular resolution of the MWA observation. 
The ASKAP repeater, FRB 20190711A, has a small positional error of $0.38$\,arcsec \citep{JP20}, well within the synthesized beam of the MWA, so we only needed to beamform the VCS data at a single position.
The other CHIME repeaters have a positional error of $\sim10$\,arcmin. Depending on the configuration of the MWA, this corresponds to $2\text{--}16$ beams for the phase II compact configuration and $\sim550$ beams for the phase I and phase II extended configuration, as shown in Table~\ref{MWAObs}.

\subsection{Data analysis}

For each of the five repeaters, we used
the VCS tied-array beamformer
to create a time series with a temporal and frequency resolution of $100\,\upmu$s and 10\,kHz at each of the selected pointing directions that cover the entire FRB positional error. 
We performed a standard FRB search over the 30.72\,MHz bandwidth and a DM range of $100\,\text{pc}\,\text{cm}^{-3}$ around the nominal DMs of the repeaters listed in Table~\ref{FRBs}. For the DM ranges searched for each repeater see Table~\ref{Results}.

\subsubsection{FRB search}\label{sec:search}

The single pulse search was performed using the {\sc presto} software package \citep{Ransom01}. Compared to other radio telescopes traditionally used for high-time resolution analysis, the MWA is generally less affected by radio-frequency interference (RFI), so we did not perform any RFI excision that is usually required at higher observing frequencies (see procedures outlined in \citealt{Swainston21}). Nonetheless, any spurious events caused by RFI can be identified from the final candidates via visual inspection.

First we dedispersed the time series using the {\sc prepdata} routine in {\sc presto}. We determined the DM search range of each observation based on the nominal DM of the target FRB (see Table~\ref{Results}). 
Specifically, we searched around the known DM with approximately $\pm50\,\text{pc}\,\text{cm}^{-3}$ but shifted that limit to within the closest multiple of $10\,\text{pc}\,\text{cm}^{-3}$, covering a DM range of $100\,\text{pc}\,\text{cm}^{-3}$. Given that there is no evidence of DM evolution for the five repeating FRBs analysed in this paper, and the largest DM variation observed for FRB 20190711A and FRB 20201124A is 7 and $10\,\text{pc}\,\text{cm}^{-3}$, respectively (likely caused by varying burst morphology; \citealt{Kumar21c, Kumar21f}), the DM range we chose is sufficient for the repeat bursts searched for here.
We adopted a DM step size of $0.1\,\text{pc}\,\text{cm}^{-3}$, resulting in 1000 DM trials for each observation. To reduce the data size, the dedispersed time series were downsampled to a lower time resolution of $400\,\upmu\text{s}$. This would not affect our search for repeating bursts given the shortest burst width is 0.64\,ms in our sample of repeating FRBs (see Table~\ref{FRBs}) and the general trend of increasing burst widths at lower frequencies (e.g. \citealt{Chawla20, Pleunis21}). 

\begin{figure}
\centering
\includegraphics[width=.5\textwidth]{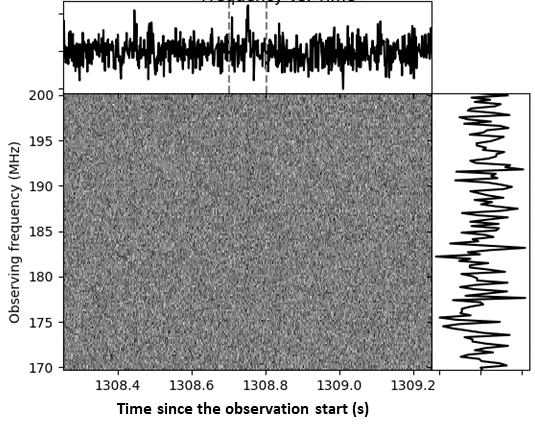}
\caption{An example dynamic spectrum of a candidate burst detected in an MWA observation (Obs ID: 1163853320) beamformed at the position of FRB 20190213A. Data have been dedispersed to a DM of $635.4\,\text{pc}\,\text{cm}^{-3}$. The dynamic spectrum has been averaged to a time resolution of 3\,ms and a frequency resolution of 0.24\,MHz. The frequency-averaged pulse profile is shown on the top panel with the candidate located between the two vertical dashed lines. 
The time-averaged spectrum is shown on the right panel.}
\label{candidate}
\end{figure}

We performed a traditional single pulse search using {\sc presto}'s {\sc single\_pulse\_search.py}, which convolves the dedispersed time series with boxcars of different widths. We chose different pulse width search ranges for the five repeating FRBs based on their brightness and the sensitivity of the MWA observations. 
While we searched up to a pulse width of 1.3\,s (the maximum scatter broadening expected at the MWA observing frequency; see Section~\ref{sec:fluence_limit} and Table~\ref{Results}) for the two brightest FRBs in our sample, i.e. FRB 20190711A and FRB 20201124A, we only searched up to 150\,ms for the other three FRBs as any signals scattered beyond 150\,ms are expected to have a flux density of $<0.2$\,Jy, $<1.4$\,Jy and $<0.2$\,Jy for FRB 20190116A, 20190117A and 20190213A, respectively (assuming a spectral index of $\alpha=-2$; see Table~\ref{Results}), which is below the MWA sensitivity of $\sim1.4$\,Jy on 150\,ms timescales.
Single pulse events detected with a S/N above six were classified as candidates (e.g. \citealt{Chawla20}, \citealt{Meyers18}, \citealt{Bannister12}). We then applied a friends-of-friends algorithm \citep{Burke11,Bannister12} to identify possible false positives. In the case of more than five individual boxcar/DM trials clustering into a candidate (e.g. \citealt{Kumar21c}), we reserved it for further inspection (see Figure~\ref{candidate} for an example candidate).

We tested our data processing and FRB detection pipeline by searching for single giant pulses from the Crab pulsar in a 154\,MHz MWA observation listed in Table~\ref{MWAObs} (Obs ID: 1165246816).  
We ran our FRB search pipeline over a chosen DM range of $56.77\pm5\,\text{pc}\,\text{cm}^{-3}$ (from the ATNF Pulsar Catalogue\footnote{\url{https://www.atnf.csiro.au/research/pulsar/psrcat/}}; \citealt{Manchester05}).
We successfully detected 89 single pulses with $\text{S/N}>10$ from this 15\,min observation (see Appendix~\ref{appendix:Crab} and Figure~\ref{Crab_spec} for the dedispersed dynamic spectrum of the brightest Crab pulse). 
In order to calculate the expected giant pulse rate from the Crab at the observing frequency of 154\,MHz, we took the Crab pulse rate of $0.35\,\text{s}^{-1}$ at 185\,MHz for a fluence $>520$\,Jy\,ms, the spectral index of $-0.7\pm1.4$ and the power-law fluence distribution $N(>F_\nu)\propto F_\nu^{-\beta}$ with $\beta=2.88\pm0.12$, all calculated by \citet{Meyers17}. We obtained a pulse rate of $0.03\text{--}0.17\,\text{s}^{-1}$ for a fluence $>1000$\,Jy\,ms (corresponding to the $10\sigma$ sensitivity of the tested observation).
Therefore, the expected number of detectable giant pulses from the Crab at 154\,MHz within our 15\,min observation is 27--153, which is consistent with 
our detection of 89 pulses, and thus verifying the utility of 
our VCS data processing and FRB detection pipeline.  

\begin{table*}
\begin{threeparttable}
\centering
\resizebox{2\columnwidth}{!}{\hspace{-0cm}\begin{tabular}{c c c c c c c c c c}
\hline
 FRB & Obs. & Expected & DM search range & Bounds of & Bounds of & Fluence upper & \multicolumn{3}{c}{Fluence expected (Jy\,ms)\tnote{6}} \\
 \cline{8-10}
 & (hr)\tnote{1} & bursts\tnote{2} & ($\text{pc}\,\text{cm}^{-3}$) &scattering (ms)\tnote{3}  & pulse widths (ms)\tnote{4} & limit (Jy\,ms)\tnote{5} & $\alpha=-1$ & $\alpha=-1.53$ & $\alpha=-2$ \\
\hline
20190116A & 2 & 0.5 & 400--500 &0.04--1217  & 5.97--1217 &  42--648 &  2--8&4--15 & 7--25 \\
20190117A & 6.8 & 1.8 & 350--450&0.22--819& 5.20--819 &  32--534 &  16--50 &30--106& 53--208 \\
20190213A & 4.2 & 0.5 & 600--700& 0.15--442  & 9.43--442 & 48--441 &  2--10 &3--18& 4--32 \\
20190711A & 9.2 & / & 550--650& 0.27--1106  & 5.06--1106 & 32--1175 &  8--276 &21--837& 47--2240 \\
20201124A & 1.9 & 30 & 350--450&4.95--1228   & 9.52--1229 & 36--488 &  9--456 &18--978& 43--1924 \\ \hline
\end{tabular}}
\caption{Derived burst properties of the five repeaters at MWA observing frequencies. \\
1: The total exposure time of each repeater with MWA; \\
2: The expected number of bursts assuming a constant burst rate extending to the MWA observing frequencies (see Table~\ref{FRBs});\\
3: The scattering timescale at the MWA observing frequency inferred from the observed scattering at higher frequencies and the contribution from the Milky Way (see Section~\ref{sec:sensi});\\
4: The pulse width at the MWA observing frequency derived by taking into account the pulse broadening due to dispersive smearing and pulse scattering (see Section~\ref{sec:fluence_limit});\\
5: The fluence upper limit for each repeater derived from Equation~\ref{eq:fluence} using the range of possible pulse widths and the flux density limits listed in Table~\ref{MWAObs}
(see Section~\ref{sec:fluence_limit});\\
6: The range of expected fluence values of known repeater bursts if 
extrapolated from their discovery frequency to an MWA observing frequency of 185\,MHz assuming a power-law spectrum with different spectral indices (see Section~\ref{sec:spectrum}).
}
\label{Results}
\end{threeparttable}
\end{table*}

\subsubsection{Determination of system sensitivity}\label{sec:sensi}

Corresponding to the $6\sigma$ threshold on S/N, we determined a flux density upper limit for each of the selected MWA observations using the radiometer equation

\begin{equation}
    S_{\text{min}} = (\text{S}/\text{N})\times \frac{\text{SEFD}}{\sqrt{n_{\text{p}}t_{\text{int}}B_\text{obs}}},
    \label{eq:sensi}
\end{equation}

\noindent where $n_{\text{p}}$ is the number of polarisations sampled, $t_{\text{int}}$ is the integration time in units of $\upmu$s, and $B_\text{obs}$ is the observing bandwidth in units of MHz (see e.g. \citealt{Meyers17}).
The overall system equivalent flux density (SEFD) is given by

\begin{equation}
    \text{SEFD} = \frac{\eta T_{\text{ant}}+(1-\eta)T_{\text{amb}}+T_{\text{rec}}}{G},
\end{equation}

\noindent where $\eta$ is the direction and frequency dependent radiation efficiency of the MWA array, $T_{\text{ant}}$, $T_{\text{amb}}$ and $T_{\text{rec}}$ represent the antenna, ambient and receiver temperatures respectively, and $G=A_\text{e}/2k_\text{B}$ is the system gain where $A_\text{e}$ is the tied-array effective area and $k_\text{B}$ is Boltzmann constant (e.g. \citealt{Lorimer04}). The radiation efficiency $\eta$ at the positions of FRBs can be calculated using a power wave based framework \citep{Ung19}; the receiver temperature is well characterised across the MWA band; and the ambient temperature is calculated from the metadata of the observation. 
To calculate the antenna temperature and gain, we need a good knowledge of the tied-array synthesised beam pattern, i.e. the product of the array factor and an individual MWA tile power pattern, which can be obtained from the phase information that points the telescope to a target position and a simulation of the tile pattern as described in \citet{Sutinjo15}. Assuming a sky temperature map at the observing frequency based on the global sky model of \citet{Costa08}, we estimated the antenna temperature and the tied-array gain by convolving the map with the tied-array beam pattern (e.g. \citealt{Sokolowski15}). For the coherent beamforming of the full (128 tiles) MWA, the SEFD is typically $\sim10^3$\,Jy at 154\,MHz \citep{Meyers17}.

Bandwidth considerations are needed for the final determination of the minimum detectable flux density in each of the selected MWA observations. As we flagged 16 of the 128 fine channels, the effective bandwidth is reduced to 87.5\% of the full 30.72\,MHz. To correct for this, we need to apply a scaling factor of $0.875^{-1/2}\approx1.07$ when converting to flux density limits. In the case of bad tiles flagged during the calibration process (see Section~\ref{sec:cal}), we also need to correct for the corresponding sensitivity loss by assuming the sensitivity scales with the number of tiles. 
Taking into account the above considerations, we arrived at a $6\sigma$ flux density upper limit for each observation, as shown in Table~\ref{MWAObs}. Note that this flux limit is based on the telescope pointing and the source location in the primary beam (with an elevation in the MWA field of $>30^\circ$ as shown in Table~\ref{MWAObs}).

\subsubsection{Pulse width and fluence estimates}\label{sec:fluence_limit}

As described above, the sensitivity of MWA observations is characterised by flux density. However, in the context of FRBs it is common to use fluence to represent their strengths. We can convert the flux density limit (which scales as $t^{-1/2}$) to a fluence limit using

\begin{equation}
    F=S_\text{min}\times(w_{\text{obs}}/1\,\text{ms})^{1/2}\,\text{Jy\,ms},
    \label{eq:fluence}
\end{equation}

\noindent which is dependent on the pulse duration ($w_{\text{obs}}$; e.g. \citealt{Hashimoto20}). The pulse durations of the five repeating FRBs have been measured at their observed frequencies, as shown in Table~\ref{FRBs}. We can estimate the expected pulse duration at the MWA observing frequency by

\begin{equation}
    w_{\nu_\text{MWA}}=\sqrt{w_{\text{obs}}^2+w_{\text{disp}}^2+w_{\text{scatter}}^2}
    \label{eq:width}
\end{equation}

\noindent where $w_{\text{obs}}$ is the observed pulse duration at higher frequencies, and $w_{\text{disp}}$ and $w_{\text{scatter}}$ account for the pulse broadening due to dispersive smearing and pulse scattering at the MWA observing frequency. 
Note that downward drifting of subpulses, which could also broaden pulse widths, is not considered here.
The dispersion smearing across the MWA channel, $\Delta\nu=10$\,kHz, can be calculated at the nominal DM values listed in Table~\ref{FRBs} for the five repeaters using (e.g. \citealt{AndersonM18})

\begin{equation}
    w_{\text{disp}}=8.3\times10^3\times\left(\frac{\Delta\nu}{\text{MHz}}\right)\times\left(\frac{\nu}{\text{MHz}}\right)^{-3}\times\left(\frac{\text{DM}}{\text{pc}\,\text{cm}^{-3}}\right)\,\text{s}.
\end{equation}

\noindent For a typical observing frequency of the MWA, $\nu=185$\,MHz, the corresponding dispersion smearing is $\sim5$\,ms.

FRBs also experience scattering. However, given the unknown environment of FRBs the scattering timescale is largely uncertain. We may infer the lower limit of the scattering timescale based on the Galactic contribution. Using the NE2001 model of electron density fluctuations \citep{Cordes02}, we obtained a minimum scattering timescale at 1\,GHz of $4.3\times10^{-5}$\,ms, $2.6\times10^{-4}$\,ms, $1.7\times10^{-4}$\,ms, $3.2\times10^{-4}$\,ms and $5.8\times10^{-3}$\,ms for FRB 20190116A, 20190117A, 20190213A, 20190711A and 20201124A, respectively. The upper limits on scattering timescales have been derived for FRB 20190116A, 20190117A, 20190213A and 20201124A to be $<11$\,ms, $<7.4$\,ms, $<4.0$\,ms and $<11.1$\,ms at 600\,MHz \citep{CHIME19b, CHIME20c, Marthi21}. Given FRB 20190711A has no published scattering measurement due to its complex time-domain structure \citep{Day20} or low S/N \citep{Kumar21c}, we assumed a typical scattering timescale of $<10$\,ms \citep{CHIME21}. 
Assuming the scattering timescale depends on frequency as $\nu^{-4}$ \citep{Bhat04}, we can obtain a range of scattering timescales at the MWA observing frequency. The final lower and upper limits on the scatter broadening can be found in Table~\ref{Results}. Note that the final range in Table~\ref{Results} was calculated individually for each observation before being combined for each repeater.

With the estimated dispersion smearing and scattering timescales, we can derive a range of pulse widths expected for each observation using Eq.~\ref{eq:width}, which can be further used to derive the final range of fluence limits for each observation using Eq.~\ref{eq:fluence}. 
The ranges of fluence limits derived from different observations covering the same repeating FRB were combined to obtain the final range of fluence limits for the repeater, as shown in Table~\ref{Results}.

\section{Results}

We performed FRB searches on 25 VCS observations at the positions of five repeating FRBs, of which four were discovered by CHIME and one by ASKAP. 
Following the automated pulse search, we further filtered all candidates with a S/N above $6\sigma$ using the friends-of-friends algorithm. The number of remaining candidates at the position of each repeater is listed in Table~\ref{MWAObs}, totalling 543 candidates discovered in 23.3\,hr of VCS data.
We visually inspected their pulse profiles and dynamic spectra, and found no evidence of a dispersion sweep, suggesting they are not real signals. An example dynamic spectrum of a candidate burst detected in an MWA observation (Obs ID: 1163853320) with $\text{S/N}=6.59$ is shown in Figure~\ref{candidate}.

We derived a flux density limit for each of the 25 MWA VCS observations searched for FRB signals, as shown in Table~\ref{MWAObs}. 
For FRB 20201124A, the flux density limits ranged between 18.41--22\,Jy. We also inferred a range of pulse widths of 9.52--1229\,ms at the MWA observing frequency of 185\,MHz for this repeater. With the derived flux density limits and pulse widths, we finally arrived at a range of fluence limits of 36--488\,Jy\,ms for FRB 20201124A.
We did a similar analysis for the remaining FRBs, and present all results in Table~\ref{Results}. Note that the three faint repeaters in our sample, i.e. FRB 20190116A, 20190117A and 20190213A, were only searched up to a pulse width of 150\,ms (see Section~\ref{sec:search}).
The uncertainties in the final fluence limits span more than an order of magnitude, which can be attributed to the largely uncertain scattering timescales. In the best case, i.e. minimal scattering, we would have detected any bursts with a fluence $\gtrsim50$\,Jy\,ms. Note that the low-frequency bursts detected by LOFAR from FRB 20180916B can reach a fluence of $308\pm10$\,Jy\,ms \citep{Pleunis21}, much higher than this fluence threshold (see Section~\ref{sec:prospect}).

In summary, no bursts from the five repeaters were detected in MWA observations. Based on their known properties, we inferred a range of fluence upper limits for each repeater, as shown in Table~\ref{Results}, which can be used to constrain the burst properties of the repeating FRB population.

\section{Discussion}

We explore implications of the above results for the five repeating FRBs analysed in this paper.
Given their burst rates estimated by CHIME/ASKAP observations (see Table~\ref{FRBs}), we calculated the expected number of bursts during the MWA observations assuming the burst rate is constant and frequency independent, as shown in Table~\ref{Results}. Based on this, there is a non-negligible chance of detecting bursts during the MWA observations of these repeaters, especially for FRB 20201124A, which can potentially emit 30 bursts within 1.9\,hr \citep{Lanman22}. Therefore, the fact that we did not detect any bursts could suggest that our assumption of a constant burst rate may be incorrect, the FRB emission has a shallow (broadband) or highly peaked (narrowband) spectrum, or the circumburst environment prevents the low frequency radio emission from escaping. In this section, we discuss the constraints placed on the properties of these repeaters by our non-detections. 

\subsection{Burst rate}\label{sec:rate}

The first explanation for our non-detection may be that the repeaters were not active during the MWA observations. If that is the case, we can further constrain the burst rates of the five repeaters at the MWA observing frequency (185\,MHz) to $<0.50\,\text{hr}^{-1}$, $<0.15\,\text{hr}^{-1}$, $<0.24\,\text{hr}^{-1}$, $<0.11\,\text{hr}^{-1}$ and $<0.53\,\text{hr}^{-1}$ for FRB 20190116A, 20190117A, 20190213A, 20190711A and 20201124A respectively, assuming the rate is not changing with time. Interestingly, the most stringent constraint on the burst rate of FRB 20190711A, $<0.11\,\text{hr}^{-1}$, is comparable to the lowest observed rate for a CHIME repeater ($0.05\,\text{hr}^{-1}$ for FRB 20190212A; \citealt{Fonseca20}). Note that the burst rates of FRB 20190711A and 20190212A are measured at different frequencies (154--215\,MHz for MWA and 400--800\,MHz for CHIME), and there seems to be an increasing trend in burst rates towards lower frequencies though it is not conclusive (e.g. \citealt{Pearlman20, Pleunis21}).

There is also another possibility that the repeaters were in quiescence during the MWA observations. That is, the five repeaters may have variable burst rates. 
This has been shown to be the case for FRB 20201124A. While this FRB has been observed to emit 48 bursts within three hours \citep{Marthi21}, its rate prior to discovery is as low as $<3.4\,\text{day}^{-1}$ (at the $3\sigma$ level assuming a Poisson rate; \citealt{Lanman22}). 
Despite the uncertainties associated with repeater activity, it is worth considering that the repeaters studied here emitted signals during the MWA observations in order to investigate the constraints our observations place on their spectral properties (see Section~\ref{sec:spectrum}) or emission environments (see Section~\ref{sec:free-free}).

\subsection{Spectral properties}\label{sec:spectrum}

Here we consider that our non-detections of FRB emission is due to low fluences 
in the MWA observing band. 
In the case of broadband FRB emission, we can constrain the spectral index using the fluence upper limits we derived for the five repeaters and their observed fluences at higher frequencies. The mean spectral index of FRB emission has been determined to be $-1.5^{+0.2}_{-0.3}$ from the summed power of 23 ASKAP FRBs over a frequency range of 1.1--1.5\,GHz \citep{Macquart19} and $-1.53^{+0.29}_{-0.19}$ by a simulation on a sample of 82 FRBs detected by Parkes, ASKAP, CHIME and UTMOST over 0.4--1.5\,GHz \citep{Bhattacharyya21}. Here we extrapolated the fluences observed at higher frequencies to the MWA observing frequency of 185\,MHz for three different spectral index values of -1, -1.53 and -2 (the same range of spectral indices explored by \citealt{Sokolowski18}) and include them in Table~\ref{Results} where they can be compared
to the fluence upper limits derived from our data.
While low frequency bursts from FRB 20190116A and 20190213A were likely undetectable in the searched MWA datasets based on the assumed mean spectral index, bursts from the other repeaters may well be detectable if they were active during the MWA observations. Although the scatter broadening of pulses can significantly reduce our sensitivity to FRB emission (as demonstrated by the large range of fluence upper limits shown in Table~\ref{Results}), our non-detection of low frequency bursts from the two brightest repeaters FRB 20190711A and 20201124A cannot be explained by scattering alone unless their spectral indices are $\alpha\gtrsim-1$, which is consistent with the conclusion in \citet{Sokolowski18}.


There is another possibility that the spectrum of repeating FRBs needs to be generalised beyond a simple power law. This could be attributed to a low instantaneous bandwidth as has been observed for FRB 20121102A, 20180916B and 20190711A with a typical fractional bandwidth of $\sim20\%$ \citep{Law17, Gourdji19, Majid20, Kumar21c} and/or clustering of spectra of repeat bursts as has been observed for FRB 20121102A (peaking around 1650\,MHz) and 20201124A (peaking around 650 and 500\,MHz; \citealt{Aggarwal21, Lanman22}). In the former case, the narrowband feature is similar to that observed for giant pulses from some pulsars (e.g. \citealt{Geyer21, Thulasiram21}), and might hint at similarities in emission mechanisms between repeating FRBs and giant pulses.

\subsection{Free-free absorption}\label{sec:free-free}

A low-frequency break in the spectrum of FRB emission is a plausible explanation for our non-detection. It has been suggested that free-free absorption by electrons in the circumburst environment can suppress low-frequency emission \citep{Ravi19b, Rajwade20b}. Here we derive an upper limit on the size of a nebula that the repeater may be embedded in in the context of free-free absorption (e.g. \citealt{Sokolowski18}).

The optical depth due to free-free absorption is given by

\begin{equation}
\begin{split}
 \tau_{\text{ff}}= & 1.1\times10^{-5}\times\left(\frac{T}{10^4\text{K}}\right)^{-1.35}\times\left(\frac{\nu}{185\,\text{MHz}}\right)^{-2.1} \\ & \times\frac{\text{DM}_{\text{ex}}^2}{f_{\text{eff}}\,L_{\text{pc}}},
\end{split}
\end{equation}

\noindent where $T$ is the electron temperature, $\nu$ is the observing frequency, $L_{\text{pc}}$ is the size of the absorbing material in pc, $f_{\text{eff}}$ accounts for the volume filling correction, and $\text{DM}_{\text{ex}}$ is the DM contribution from the absorbing material in $\text{pc}\,\text{cm}^{-3}$ \citep{Condon16}. $\text{DM}_{\text{ex}}$ can be estimated by the DM values in excess of the Milky Way contributions in the directions of FRBs \citep{Cordes02} and a further Milky Way halo contribution of $15\,\text{pc}\,\text{cm}^{-3}$ \citep{Shannon18}. 

If free-free absorption is responsible for the absence of low frequency emission, we can place a constraint on the size of the absorbing medium by requiring $\tau_{\text{ff}}>\text{ln}(F_{\text{exp}}/F_{\text{lim}})$, where $F_{\text{exp}}$ is the expected fluence at the observing frequency for a spectral index of $-1.53^{+0.29}_{-0.19}$ \citep{Bhattacharyya21} and $F_{\text{lim}}$ is the fluence upper limit derived from the MWA observation (see Table~\ref{Results}). In the case of minimum scattering, we find $f_{\text{eff}}\,L_{\text{pc}}<1.00\times(T/10^4\text{K})^{-1.35}\,\text{pc}$ (using $F_{\text{exp}}=106$\,Jy\,ms), $0.92\times(T/10^4\text{K})^{-1.35}\,\text{pc}$ (using $F_{\text{exp}}=837$\,Jy\,ms) and $[0.22\text{--}2.50]\times(T/10^4\text{K})^{-1.35}\,\text{pc}$ for FRB 20190117A, 20190711A, and 20201124A, respectively. Note that the fluence upper limits on the other two repeaters are not sufficient for constraining the size of their absorbing media. 

We can compare our constraints on the size of an ionized nebula with those derived for other repeaters, such as FRB 20121102 and FRB 20180916B. The lowest frequency detection of FRB 20121102 was reported at 600\,MHz by CHIME \citep{Josephy19}. However, a search for low-frequency emission from FRB 20121102 with LOFAR resulted in a fluence upper limit of 42\,Jy\,ms at 150\,MHz \citep{Houben19}. If we adopt the DM value derived from Balmer line measurements for the DM contribution from the host galaxy, $\text{DM}_\text{host}\lesssim324\,\text{pc}\,\text{cm}^{-3}$ \citep{Tendulkar17}, the size of the ionized region surrounding FRB 20121102 can be constrained to $\lesssim2.85$\,pc, which is comparable to the constraints derived above. Different than FRB 20121102 and the repeaters studied here, FRB 20180916B was reported to have low-frequency emission between 110--188\,MHz, suggesting a lower limit on the size of an ionized nebula $\gg0.16\times(T/10^4\text{K})^{-1.35}\,\text{pc}$ \citep{Pleunis21}. If the population of repeating FRBs analysed here have a similar environment as FRB 20180916B, we would expect a nebula size of $\sim1$\,pc, comparable to the Crab Nebula. 

\subsection{Comparison of FRB searches with the MWA}

Previously, the MWA has been employed to perform FRB searches
in the image domain \citep{Tingay15, Rowlinson16, Sokolowski18}. 
Here we present the first FRB search using MWA VCS observations. Compared to the imaging mode, the VCS data features a much higher time resolution, increasing our sensitivity to short duration ($\sim$ms) FRB emission which would otherwise be diluted by the 0.5\,s coarse sampling of the standard correlator. Table~\ref{compare} displays a comparison of the fluence sensitivity of our search with previous works. Note that while \citet{Tingay15} and \citet{Rowlinson16} performed a blind search for FRB emission in the wide field of view of the MWA, \citet{Sokolowski18} specifically targeted the bright FRBs detected by ASKAP via a shadowing observing strategy. We can see the VCS observations used in this work demonstrate the best sensitivity. In the case of minimal scattering, the VCS observation is more than an order of magnitude more sensitive than the standard observation, and thus is most promising for searching for FRBs in the future.

\begin{table*}
\centering
\resizebox{1.2\columnwidth}{!}{\hspace{-0cm}\begin{tabular}{c c c c c}
\hline
 Reference & Obs. & Freq. & Time/Freq. & Sensitivity \\
 & (hr) & (MHz) &  resolution &(Jy\,ms) \\
\hline
\citet{Tingay15} & 10.5 & 156 & 2\,s\,/\,1.28\,MHz & 700\\
\citet{Rowlinson16} & 100 & 182 & 28\,s\,/\,30.72\,MHz & 7980\\
\citet{Sokolowski18} & 3.5 & 185 & 0.5\,s\,/\,1.28\,MHz & 450--6500\\
This work & 24.1 & 144--215 & $400\,\upmu$s\,/\,10\,kHz & 32--1175 \\
\hline
\end{tabular}}
\caption{Comparison between FRB searches using the MWA.
}
\label{compare}
\end{table*}

\subsection{Future prospects}\label{sec:prospect}

Considering the ongoing searches for FRBs with multiple different facilities, e.g., Parkes \citep{Keane18}, ASKAP \citep{Bannister17}, CHIME \citep{CHIME18}, UTMOST \citep{Caleb16} and FAST \citep{Jiang19}, we expect more repeating FRBs to be discovered that can be observed by the MWA in the future.
Since most FRBs are discovered by CHIME, here we focus on the prospect of CHIME repeating FRB follow-up using the MWA. At the location of MWA, we are able to observe only those FRBs with low declinations ($\lesssim30$\,deg), amounting to $\sim$20\% of the whole population based on the sky distribution of FRBs (see figure 10 in \citealt{CHIME21}). Therefore, given that 18 repeating FRBs have been discovered by CHIME during its first year of operation \citep{CHIME19b, CHIME20c}, we expect to be able to follow up $\sim4$ new repeating FRBs using the MWA per year assuming that the discovery rate is constant. Note that we would ideally target repeating FRBs that were away from the Galactic plane where the sky temperature is much higher, as well as choose those that show lower levels of scattering. 

\begin{figure}
\centering
\includegraphics[width=.5\textwidth]{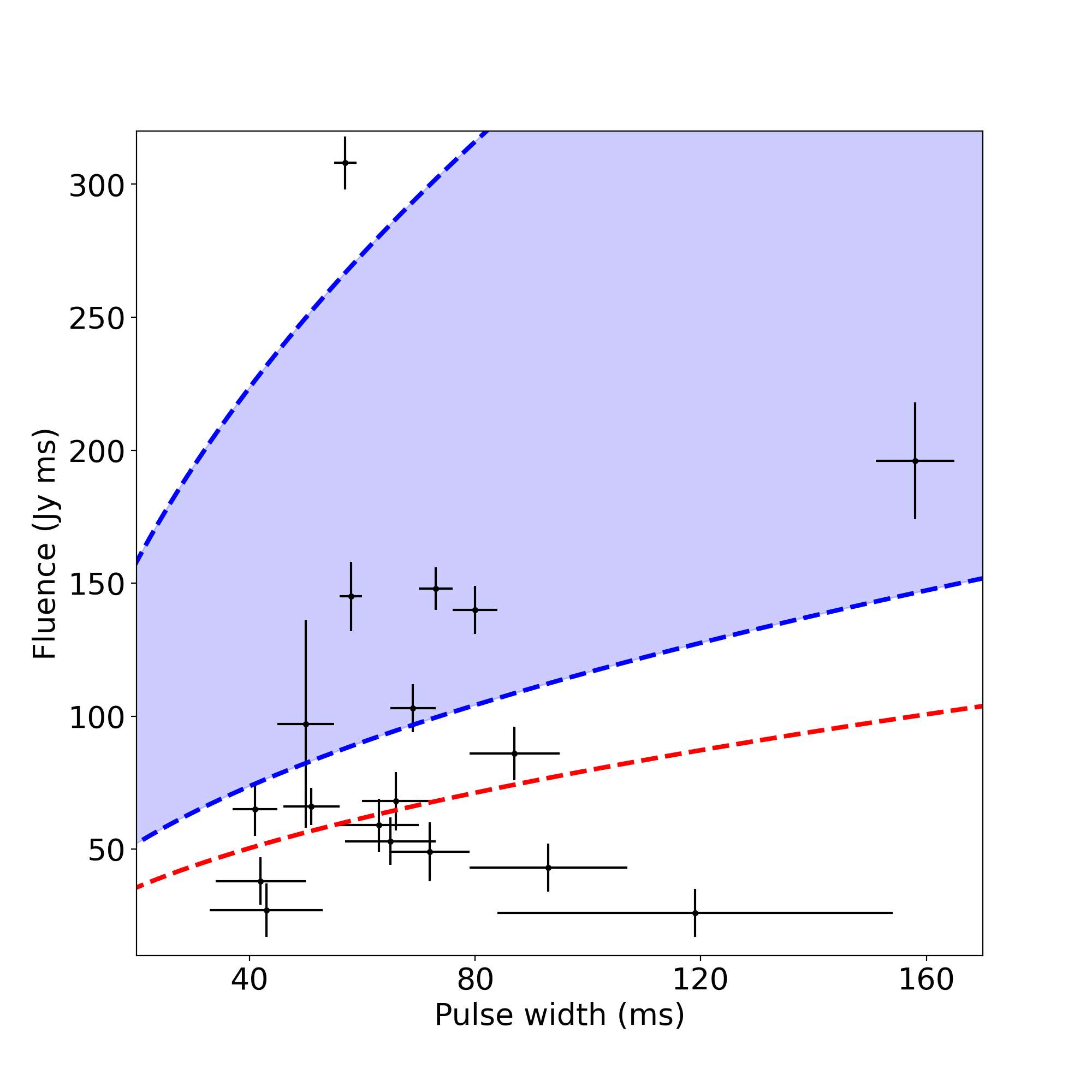}
\caption{Fluences of the bursts reported by \citet{Pleunis21} from the repeating FRB 20180916B at 110--188\,MHz versus their pulse widths. The black points represent the 18 bursts detected by LOFAR, the blue region represents the range of fluence upper limits derived from the MWA observations analysed in this work, and the dashed red line represents a typical VCS sensitivity for near-zenith observations with the MWA.}
\label{detectability}
\end{figure}

A very interesting source among the population of repeating FRBs is FRB 20180916B, which features a periodicity of 16.3\,day and low frequency emission detected by LOFAR in the frequency range of 110\textendash188\,MHz \citep{Pleunis21}, overlapping with the MWA observing band. Note that this FRB is unobservable by the MWA due to its location in the northern hemisphere. However, the fluences of the bursts from this repeater range between 
26 to 308\,Jy\,ms and are therefore close to the fluence upper limits we derived from the MWA VCS observations studied in our analysis.
It is therefore instructive to explore the detectability of these bursts by the MWA if they appeared in the MWA field of view. In Figure~\ref{detectability}, we plot the pulse widths and fluences of the 18 detected bursts (black points) and the fluence upper limits derived from the MWA observations included in this work (blue region). Considering the sensitivity of the MWA observations in this work are limited by the low elevations of the targeted FRBs in the MWA field of view, we also plot a typical sensitivity near the zenith of the full MWA for comparison (\citealt{Meyers17, Meyers18}; dashed red line). While the observations in this work would only have been sensitive enough to detect the seven brightest bursts, a typical VCS observation near the zenith could have detected up to 12 bursts, and therefore $\sim70$\% of the bursts reported by \citet{Pleunis21} from the repeating FRB 20180916B at low frequencies ($<200$\,MHz). Therefore, if another repeating source with emission properties similar to FRB 20180916B becomes known in the MWA sky in the future, follow-up campaigns with the MWA would be capable of detecting low frequency bursts.

Apart from follow-up observations of repeaters, we plan to conduct an all-sky FRB search with the MWA. Here we consider searching all the VCS archival data for FRBs. The rate of FRB detections by the MWA can be estimated using

\begin{equation}
    \frac{R^{F1}_{\nu1}}{R^{F2}_{\nu2}}=\left(\frac{\nu1}{\nu2}\right)^{-\alpha\beta}\times\left(\frac{F1}{F2}\right)^{\beta-1},
\label{eq:rate}
\end{equation}

\noindent where $R^{F1}_{\nu1}$ ($R^{F2}_{\nu2}$) is the sky rate above a fluence of $F1$ ($F2$) at a radio frequency of $\nu1$ ($\nu2$), $\alpha$ is the average spectral index ($-1.53^{+0.29}_{-0.19}$; \citealt{Bhattacharyya21}), and $\beta$ is the power-law index parameterizing the fluence distribution ($-1.40\pm0.11^{+0.06}_{-0.09}$; \citealt{CHIME21}). Here we adopted the latest sky rate of $820\pm60^{+220}_{-200}\,\text{sky}^{-1}\,\text{day}^{-1}$ for FRBs (including both repeaters and non-repeaters) with a fluence $>5$\,Jy\,ms at 600\,MHz and a scattering time less than 10\,ms at 600\,MHz, which is based on the first large sample of FRBs \citep{CHIME21} and consistent with the rate reported from the Green Bank North Celestial Cap (GBNCC) survey in the 300--400\,MHz band \citep{Parent20}. Scaling this rate down to the MWA observing frequency of 185\,MHz and a typical VCS sensitivity of 50\,Jy\,ms as shown in Figure~\ref{detectability} using Eq.~\ref{eq:rate}, we expect a rate of $15\text{--}78\,\text{sky}^{-1}\,\text{day}^{-1}$. Given the $\sim1$\% instantaneous sky coverage of the MWA \citep{Tingay13, Tingay15}, we expect on average 1--7\,days of VCS observations would yield an FRB detection. It is noteworthy that the computational cost of processing such a large amount of data is still a challenge \citep{Trott13}.

\section{Conclusions}

In this paper, we have searched for low-frequency emission from five repeating FRBs using the VCS observations in the MWA archive. This is the first time that the MWA VCS has been used to search for bursts from repeating FRBs. The 25 MWA VCS observations analysed ranged in integration times between 15\,min and 1.5\,hr (23.3\,hr in total). 
As a result of this work, we come to the following main conclusions:

\begin{enumerate}
    \item 
    If FRB 20190711A and FRB 20201124A were active during the MWA observations, then we can constrain their spectral index to $\alpha\gtrsim-1$ assuming a broadband spectrum, shallower than the mean spectral index estimated on a sample of FRBs \citep{Bhattacharyya21}.
    \item 
    The fluence upper limits derived from the MWA observations in the case of minimum scattering on the FRB emission (contributed only by the Milky Way) enable us to constrain the size of the absorbing medium to $<1.00\times(T/10^4\text{K})^{-1.35}\,\text{pc}$, $<0.92\times(T/10^4\text{K})^{-1.35}\,\text{pc}$ and $<[0.22\text{--}2.50]\times(T/10^4\text{K})^{-1.35}\,\text{pc}$ for FRB 20190117A, 20190711A, and 20201124A respectively, which is comparable to the size limit on FRB 20121102 \citep{Tendulkar17, Houben19} and lower limit for FRB 20180916B \citep{Pleunis21}.
    \item Compared to previous MWA searches for low-frequency FRB emission using the standard correlator with a minimum temporal resolution of only 0.5\,s \citep{Tingay15, Rowlinson16, Sokolowski18}, our VCS observations with a temporal resolution of $400\,\upmu$s are more than an order of magnitude more sensitive to FRB signals except in the case of severe pulse broadening due to scattering beyond $\sim0.5$\,s.
    \item A comparison between the typical sensitivity of the MWA VCS observation at the zenith and the fluences detected by LOFAR from FRB 20180916B reveals that the MWA would be able to detect $\sim70$\% of the low frequency bursts from a repeater like FRB 20180916B.
\end{enumerate}

In conclusion, our non-detections are likely due to the limited total integration time of our VCS observations, during which the repeaters were likely inactive. However, we cannot rule out the possibility of a shallow spectrum or a dense environment. In order to detect the low-frequency counterparts of repeating FRBs or clearly constrain their burst properties, we need a combination of the reanalysis of archival VCS observations, which will continue to be collected over the duration of the SMART survey, and potentially dedicated VCS observing campaigns.
Higher frequency instruments such as CHIME and ASKAP can be used to determine the active windows of repeating FRBs when targeted observations with the MWA will be most useful.
Currently, the high time resolution dataset that covers a large part of the Southern Sky, which is only made possible via access to high time resolution data archives, presents a unique opportunity for transient science. Given this rich dataset, a blind search is a logical and efficient way of identifying low frequency FRBs and repeaters.

\section*{Acknowledgements}

We thank the referee for their constructive feedback, which has improved the paper.

This scientific work uses data obtained from Inyarrimanha Ilgari Bundara / the Murchison Radio-astronomy Observatory, operated by CSIRO. We acknowledge the Wajarri Yamatji People as the Traditional Owners and native title holders of the Observatory site. Support for the operation of the MWA is provided by the Australian Government (NCRIS), under a contract to Curtin University administered by Astronomy Australia Limited.

This work was supported by resources provided by the Pawsey Supercomputing Centre with funding from the Australian Government and the Government of Western Australia. We acknowledge use of the CHIME/FRB Public Database, provided at https://www.chime-frb.ca/ by the CHIME/FRB Collaboration.

GEA is the recipient of an Australian Research Council Discovery Early Career Researcher Award (project number DE180100346) funded by the Australian Government. 

The following software and packages were used to support this work:
{\sc Astropy} \citep{TheAstropyCollaboration2013,TheAstropyCollaboration2018}, 
{\sc numpy} \citep{vanderWalt_numpy_2011}, 
{\sc scipy} \citep{Jones_scipy_2001}, 
{\sc matplotlib} \citep{hunter07}, {\sc presto} \citep{Ransom01},   docker\footnote{\href{https://www.docker.com/}{https://www.docker.com/}}, 
singularity \citep{kurtzer_singularity_2017}.
This research has made use of NASA's Astrophysics Data System. 

\section*{Data availability}

The MWA data underlying this paper are available in the MWA Long Term Archive at \url{http://ws.mwatelescope.org/metadata/find}. The intermediate data products will be shared on reasonable request to the corresponding author.




\bibliographystyle{mnras}
\bibliography{bib} 



\appendix

\section{The brightest pulse detected from the Crab pulsar with the MWA observation}\label{appendix:Crab}

\begin{figure}
\centering
\includegraphics[width=.5\textwidth]{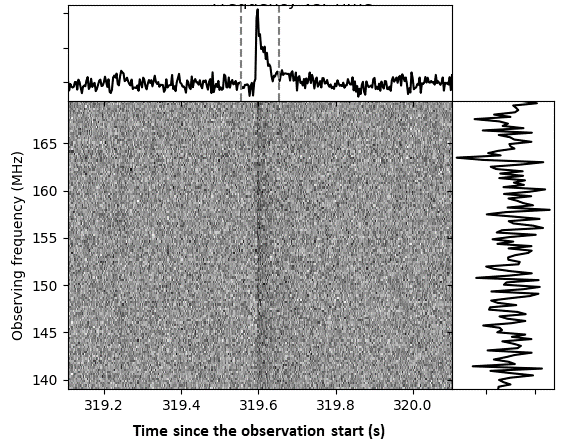}
\caption{The dynamic spectrum of the brightest pulse detected with an MWA observation from the Crab pulsar. Data have been dedispersed to a DM of $56.76\,\text{pc}\,\text{cm}^{-3}$. The dynamic spectrum has been averaged to a time resolution of 3\,ms and a frequency resolution of 0.24\,MHz. The frequency-averaged pulse profile is shown on the top panel (between the two vertical dashed lines), and the time-averaged spectrum is shown on the right panel.}
\label{Crab_spec}
\end{figure}

In Figure~\ref{Crab_spec}, we present the dynamic spectrum of the brightest pulse detected from the Crab pulsar with an MWA observation (Obs ID: 1165246816) at a DM of $56.76\,\text{pc}\,\text{cm}^{-3}$, S/N of 21.84 and time since the observation start of 319.59\,s (see Section~\ref{sec:search}).

\bsp	
\label{lastpage}

\end{document}